\documentclass{article}

\usepackage{microtype}
\usepackage{graphicx}
\usepackage{subfigure}
\usepackage{enumitem}
\usepackage{booktabs} 

\usepackage{multirow}

\usepackage{hyperref}

\usepackage[accepted]{icml2021}

\icmltitlerunning{The Golden Circle: Creating Sociotechnical Alignment in Content Moderation}

\begin{document}

\twocolumn[
\icmltitle{The Golden Circle: Creating Socio-technical Alignment in Content Moderation}



\icmlsetsymbol{equal}{*}

\begin{icmlauthorlist}
\icmlauthor{Abhishek Gupta}{msft}
\icmlauthor{Iga Kozlowska}{fb}
\icmlauthor{Nga Than}{cuny}
\end{icmlauthorlist}

\icmlaffiliation{msft}{Montreal AI Ethics Institute \& Microsoft, Montreal, Canada}
\icmlaffiliation{fb}{Meta, Seattle, USA}
\icmlaffiliation{cuny}{City University of New York - The Graduate Center, New York, USA}

\icmlcorrespondingauthor{Abhishek Gupta}{abhishek@montrealethics.ai}


\vskip 0.3in
]



\printAffiliationsAndNotice{\icmlEqualContribution} 

\begin{abstract}
This paper outlines a conceptual framework titled \textit{The Golden Circle} that describes the roles of actors at individual, organizational, and societal levels, and their dynamics in the content moderation ecosystem. Centering “harm-reduction” and “context-moderation,” it argues that the ML community must attend to multi-modal content moderation solutions, align their work with their organizations’ goals and values, and pay attention to the ever changing social contexts in which their sociotechnical systems are embedded. This is done by accounting for the why, how, and what of content moderation from a sociological and technical lens. 
\end{abstract}

\section{Introduction}
\label{submission}

Content moderation has become an increasingly challenging sociotechnical issue as sociolinguistic norms change at an ever-increasing rate and more people than engage online. While social scientists have focused on social issues of content moderation like poor labor conditions \cite{roberts2019behind} \cite{sablosky2021dangerous}, technologists have mainly focused on developing and perfecting AI solutions to remove harmful content from the information ecosystem \cite{vijayaraghavan2019interpretable}. Moving beyond the dichotomy of the technical vs. the social, we develop \textit{the Golden Circle Framework} (GCF), an interdisciplinary approach to tackle the content moderation problem, which takes into account both individual ML researchers, developers' positions, and the collective action of organizations and other society actors.

\section{The Golden Circle}

A growing body of literature has pointed out that one should move beyond content moderation to “context-moderation” and “harm-reduction” \cite{caplan2018content}. The GCF, an adaptation of the Golden Circle by Simon \citet{sinek2009start}, provides a road map for evaluating content moderation harms and organizing responses to mitigate them on an individual, organizational, and societal level keeping in mind both the technical and sociological interventions. The central tenet of the GCF is that these vectors of effort (having both direction and magnitude as per the notion in Physics) need to be aligned to make content moderation efforts effective.
 \vspace*{-.4cm}
\begin{figure}[htp]
    \centering
    \includegraphics[width=5.5cm]{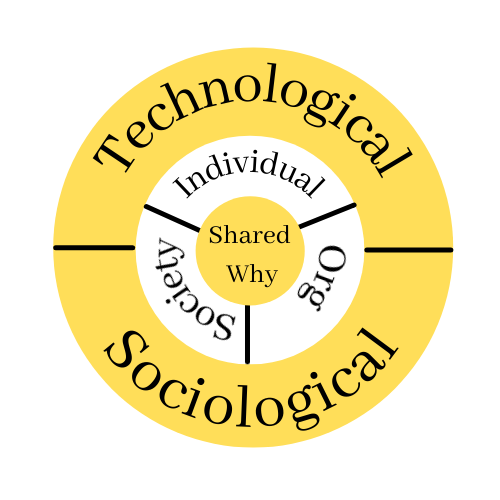}
    \caption{The Golden Circle Framework. The inner most layer represents the "Why" that all actors agree on. The second layer is the "How." The outer most layer represents the "What." }
    \label{fig:galaxy}
\end{figure}

The motivating "Why" of content moderation is maintaining social cohesion and trust that underpins a fair and healthy democracy. Mistrust incurs steep social costs by creating sharp societal division in areas like healthcare \cite{sylvia2020we} and politics \cite{hameleers2020misinformation} and is likely in the long-run unsustainable in a democratic society. Yet, scholarship on content moderation has identified numerous harms to groups and social institutions when user-generated content is not sufficiently moderated \cite{freelon2020false, schradie2019revolution}. The following harms must be reduced to preserve a healthy information ecosystem that supports social trust and cohesion.

\begin{enumerate}[label=(\alph*)]
\item \textbf{Radicalization:} the availability of radical, extremist, violent and otherwise objectionable content online may lead to the radicalization of political or cultural views, whether through organized radicalization efforts such as ISIS recruitment networks or less organized methods like the spread of misinformation and disinformation (e.g. QAnon conspiracy theories) \cite{zuckerman2019unreality, lewis2018alternative}. Radicalization and incitement to hatred or violence result from harmful content not being detected before it is posted or not being taken down after it has been posted.

\item \textbf{Polarization:} the “echo chamber” effect results in users seeing more and more content that they are already more prone to engage with, over time resulting in highly homogeneous content consumption \cite{bail2018exposure}. Polarized political worldviews result in not only high levels of mistrust of the “other” but also different versions of the truth, or different epistemologies \cite{freelon2020false}. No longer able to agree on basic facts and a common reality, the public suffers polarization that makes democratic participation, deliberation, and decision-making difficult. Polarization results not necessarily from harmful content not being removed, but the patterns of information that are systematically served to some and not others \cite{liao2014can}. 

\item \textbf{Online harassment:} online bullying, harassment, and discrimination, often of marginalized groups or individuals based on race, gender, sexual orientation, disability, religion, or ethnicity, whether targeted at specific individuals or entire groups such as Rohingya Muslims \cite{siddiquee2020portrayal}, results in exclusion, marginalization, and in the most extreme cases physical violence, and even genocide. This virtual \cite{patton2014social} and physical violence results from harmful content not being detected before it is posted or not being taken down after it has been posted.

\item \textbf{Attention control:} online content has been shown to be addictive by design. Internet “influencers” produce content that could go “viral” to get more followers; companies post “click-bait” to increase ad revenue \cite{wu2017attention}. Ubiquitous social media content production and consumption result in increased feelings of isolation, decreasing ability to focus, feelings of a sense of loss of self-control, and the opportunity cost of not engaging in more healthy or productive activities \cite{sujarwoto2019tool}. Attention deficit results from the sheer volume of content production and the commercial incentives to generate attention-grabbing and low-quality content (volume over quality). 

\item \textbf{Labor:} moderating online content incurs emotional and psychological costs on low-paid wage workers, who often reside in the developing world, to screen objectionable content \citep{roberts2019behind}. 

\item \textbf{Environmental:} Automated content moderation systems which utilize large language models require a sprawling infrastructure that uses up natural resources and vast amounts of energy to operate \cite{bender2021dangers}. They incur environmental costs that must be taken into account when deciding how to regulate content production and moderation. 
\end{enumerate}

Why have these well documented harms not been adequately addressed? Our framework shows that even though everyone may agree on the “Why” (healthy information ecosystem), the “How” and the “What” are not aligned. In other words, though few would disagree with the need to maintain social trust and a healthy information ecosystem, current solutions to content moderation are not aligned with how individuals and organizations operate, especially when it comes to their incentives. 

Our framework provides a road map for realigning content moderation solutions to the motivating “Why” while respecting the fundamental “How” of individuals and organizations that are largely immutable. The "How" for ML developers is to seek job rewards like promotions and benefits. Yet organizations don't incentivize harm reduction through improve content moderation, so sufficient technical improvements are not innovated. There is a feeback loop from organization to individual. Similarly, the "How" of organizations is to maximize profit while avoiding negative public opinion. Yet society, through civic and legislative action, has not demanded sufficient organizational change. There is a feedback loop from society to organization. Each stakeholder level feedback enables the other and must be aligned. Then, and only then, can technical and sociological solutions - the "What" - be successfully implemented. 

We propose this framework not as a single, one-time “solution” to content moderation but rather as a way to better align the vectors to drive toward outcomes that reduce harm, acknowledging that eliminating all harm completely is unrealistic. Let's turn now to the "What" - proposed solutions to reduce content moderation harms.

\renewcommand{\arraystretch}{1.7}
\begin{table*}[h]
    \centering
    \begin{tabular}{|p{2.5cm}|p{2.5cm}|p{3.2cm}|p{7.3cm}|}
         \hline
         \textbf{Why} & \textbf{Who}   & \textbf{How} & \textbf{What} \\
         \hline
Preserve social cohesion and social trust sufficient to sustain a robust and fair democracy
& \textbf{Individual:} ML developer designing content moderation algorithms &  Seek job rewards like promotion and benefits &  
\vspace{-.55cm}
\begin{itemize}  [label={},leftmargin=*]
\item \textbf{Technical:} Improve algorithms used in content moderation.  
\item \textbf{Sociological:} ML developers should recognize their own implicit biases as well as technical limitations in solving what is a complex social problem
\end{itemize}\\
 \cline{2-4}
 & \textbf{Organization:} social media platform  
& Maximize profit while avoiding negative public opinion &  \textbf{Sociological:}

\begin{itemize}[noitemsep,topsep=0pt,nosep]

  \item Increase resourcing for ethical AI work and diversify AI workforce 
  \item Build interdisciplinary teams
  \item Restructure ML developer incentives to reward human-centered design
  \item Develop business models that more fairly distributes platform benefits
  \item Foster a culture of transparency, integrity, and accountability 
\vspace{-1pc}
\end{itemize} \\

\cline{2-4}
       & \textbf{Society}  & Sustain a healthy information ecosystem through better content moderation&  \textbf{Sociological:} Hold social media platforms accountable through regulatory and policy instruments
         NGOs and academics must continue to scrutinize content moderation practices
         Sustain robust public dialogue on the role of social media in society \\
\hline

\end{tabular}
    \caption{Description of the role of each actor within the Golden Circle Framework}
    \label{tab:my_label}
\end{table*}

\section{Individual}

What can those who build algorithms that power social media experiences do about it? ML developers are uniquely positioned to offer technical solutions to sociotechnical problems, so their improvements are only a piece of the puzzle, but let us consider some proposals. 

To reduce the amount of harmful content circulating in the information ecosystem, we need to improve the accuracy of models in identifying extremist, violent, or otherwise objectionable content, whether text, image, video or more and more some combination of each. The difficulties lie in algorithms understanding cultural references and symbols that are unspecified or implicit. To understand this nuance, models need to be better able to capture the cultural context within which meaning is socially produced \cite{caplan2018content}. For example, \citet{gao2017detecting} improved logistic regression and neural net hate speech detection models by introducing contextual features like screen name, full comment thread, and the news article associated with each thread. They enlarged the object of analysis by including culturally-relevant context as features in the model.  Improved model accuracy will also have the spillover effect of minimizing as much as possible the necessity for human review of objectionable content. The unfair toll on human reviewers has been well documented \cite{roberts2019behind}. 

Other machine learning innovations can have positive indirect effects on content moderation like attempts at reducing the echo chamber effect. Recommendation algorithms need to be redesigned to optimize for a healthy plurality of content \cite{sheth2011towards} rather than the spiral of homogeneous content that is the product of algorithms that optimize chiefly for views and clicks. Similarly, these algorithms need to consider features that can help reduce the addictiveness of content, focusing less on virality more on quality content \cite{delOlmo2008evaluation}. This can also help reduce the incentives to serve customers more and more radical content instead privileging more balanced and higher quality content. Finally, the environmental impacts of running large language models used in content production and moderation have been well documented. Algorithmic efficiency up and down the whole production chain must be improved or new solutions introduced to reduce environmental burden \cite{lacoste2019quantifying}.

Recognizing that ML developers have a bias for technical solutions, next we need to expand the context within which models get built. This requires examining the organizational structures of social media platforms.

\section{Organization}

How can more research like what discussed above and additional innovations be encouraged and incentivized within social platform organizations? Currently there is a misalignment between organizational goals and culture, incentivizing ML teams to build bigger and better language models \cite{chelba2012large} that are better at targeting content to users, not necessarily reducing harmful content or otherwise improving content quality. 
Unfortunately, there is limited available research on the kinds of organizational processes, tools or cultures that are most practicable and conducive to more ethical algorithmic solutions \cite{baucus2005designing}. Some innovative social scientific research, however, is moving in that direction. \citet{moss2020ethics} published a report on “ethics owners,” those in tech organizations assigned to address ethical concerns,  champion ethical causes, designs, development, and deployment of technology from within the tech industry. Central to our argument, they find that the “personal ethics of the ethics owner do not always align with those of the corporation” and that organization practices and business models “have implications for how ethics owners approach their work.” 

Our framework proposes several possibilities by which social media platforms can align content moderation practices to more ethical outcomes while still maintaining profitability. These include increasing resources for ethical AI work and diversifying the AI workforce. An extensive body of research shows that more diverse teams build better products, increase sales revenue, leading to better organizational outcomes \cite{page2008difference, page2019diversity, muller2019learning, harrington2010pop, jehn1999differences,herring2009does}. Investing in these areas should not cut into the bottom line. Second, organizations need to build interdisciplinary teams composed of technical and non-technical roles so that social implications of poor content moderation can be brought to the fore. This does not require hiring external personnel, but rather expanding the kind of experts that have a seat at the table. Third, organizations must restructure ML developer incentives to reward the development of algorithms that put human well-being at the center. This can include alternative ways of measuring employee progress or impact by de-emphasizing speed and quantity and instead focusing on harm reduction, alignment with the latest social scientific research, and customer satisfaction. Fourth, organizations should explore business models that more fairly distributes platform benefits. This could include compensation for content generators or data subjects while incentivizing quality content \cite{wohn2019volunteer}. Technology companies routinely update, tweak, or “pivot,” and sometimes completely invent new business models \cite{shestakofsky2017working, ravenelle2019hustle, griesbach2019algorithmic}. Just because the current advertising-based business model of social media platforms is the most widespread and profitable does not mean it is the only viable model possible. Investing in business model innovation will create incentives for model developers to think more creatively about how machine learning can be employed for different purposes than it currently is. Finally, organizations must foster a culture of transparency, integrity, and accountability where critical research and criticism of company policy is not only tolerated but encouraged. ML researchers and developers cannot propose innovative solutions without first being able to point to what is not working. 

\section{Society}

Broadening our lens even further, and to force the kind of realignments mentioned above, here we propose some actions for legislators, NGOs and academe. 

First, legislators and regulators must limit the outsized power social media platforms currently hold on content moderation \cite{langvardt2017regulating}. Much like there are federally-mandated standards for content production for other industries like film, television, journalism and other media, so too we need democratically-elected representatives to set the standard for social media. Second, legislators must consider laws that incentivize platforms to do better content moderation by developing legal and regulatory instruments to hold them accountable for social harms when they refuse or fail to comply \cite{gorwa2020algorithmic}. Third, NGOs and academics must continue to scrutinize content moderation practices, centering especially those that are most harmed by failures, to help organizations see problems they might not otherwise see, while holding them accountable for fixing them. Due to corporate secrecy and legal protections around proprietary information and trade secrets, content moderation practices continue to remain black boxes, even if some efforts at transparency reports have improved that marginally in recent years. This is why support from legislators is needed to make it easier for civil society organizations and academics to study and evaluate content moderation practices by mandating certain levels of transparency in the name of public interest. Finally, citizens, social rights organizations, journalists, academics, and policymakers all need to participate in a robust dialogue on the role of social media in society and negotiate a set of common principles that can guide policy action. The technical and complex nature of machine learning pose challenges for this kind of informed conversation, but as the field of machine learning becomes more and more mainstream so too its broader understanding will improve. 

\section{Conclusion}

To conclude, efforts like ours and many others \cite{sloane2019ai, lindgren2020social} at bridging the gap between machine learning experts and social science researchers (and the broader public) strive to better align the practices of all stakeholders (ML developers, organizations and society) for the purpose of generating improved social outcomes. Our framework points out current day misalignments so that we can address these gaps in a more organized manner. This is needed because the harms of an unhealthy information ecosystem are real, as evidenced by recent events like the attack on the U.S. Capitol. Our framework uses the case study of content moderation as one field within which ML solutions are applied. But it has significance for other sociotechnical AI-based problems as ML becomes applied to more and more aspects of our social lives. 

\nocite{langley00}

\bibliography{Golden_Circle}

\begin{thebibliography}{39}
\providecommand{\natexlab}[1]{#1}
\providecommand{\url}[1]{\texttt{#1}}
\expandafter\ifx\csname urlstyle\endcsname\relax
  \providecommand{\doi}[1]{doi: #1}\else
  \providecommand{\doi}{doi: \begingroup \urlstyle{rm}\Url}\fi

\bibitem[Bail et~al.(2018)Bail, Argyle, Brown, Bumpus, Chen, Hunzaker, Lee,
  Mann, Merhout, and Volfovsky]{bail2018exposure}
Bail, C.~A., Argyle, L.~P., Brown, T.~W., Bumpus, J.~P., Chen, H., Hunzaker,
  M.~F., Lee, J., Mann, M., Merhout, F., and Volfovsky, A.
\newblock Exposure to opposing views on social media can increase political
  polarization.
\newblock \emph{Proceedings of the National Academy of Sciences}, 115\penalty0
  (37):\penalty0 9216--9221, 2018.

\bibitem[Baucus \& Beck-Dudley(2005)Baucus and
  Beck-Dudley]{baucus2005designing}
Baucus, M.~S. and Beck-Dudley, C.~L.
\newblock Designing ethical organizations: Avoiding the long-term negative
  effects of rewards and punishments.
\newblock \emph{Journal of Business Ethics}, 56\penalty0 (4):\penalty0
  355--370, 2005.

\bibitem[Bender et~al.(2021)Bender, Gebru, McMillan-Major, and
  Shmitchell]{bender2021dangers}
Bender, E.~M., Gebru, T., McMillan-Major, A., and Shmitchell, S.
\newblock On the dangers of stochastic parrots: Can language models be too big?
\newblock \emph{Proceedings of FAccT}, 2021.

\bibitem[Caplan(2018)]{caplan2018content}
Caplan, R.
\newblock Content or context moderation? artisanal, community-reliant, and
  industrial approaches.
\newblock \emph{Data \& Society. https://datasociety.
  net/library/content-or-context-moderation}, 2018.

\bibitem[Chelba et~al.(2012)Chelba, Bikel, Shugrina, Nguyen, and
  Kumar]{chelba2012large}
Chelba, C., Bikel, D., Shugrina, M., Nguyen, P., and Kumar, S.
\newblock Large scale language modeling in automatic speech recognition.
\newblock \emph{arXiv preprint arXiv:1210.8440}, 2012.

\bibitem[Del~Olmo \& Gaudioso(2008)Del~Olmo and
  Gaudioso]{delOlmo2008evaluation}
Del~Olmo, F.~H. and Gaudioso, E.
\newblock Evaluation of recommender systems: A new approach.
\newblock \emph{Expert Systems with Applications}, 35\penalty0 (3):\penalty0
  790--804, 2008.

\bibitem[Freelon et~al.(2020)Freelon, Marwick, and Kreiss]{freelon2020false}
Freelon, D., Marwick, A., and Kreiss, D.
\newblock False equivalencies: Online activism from left to right.
\newblock \emph{Science}, 369\penalty0 (6508):\penalty0 1197--1201, 2020.

\bibitem[Gao \& Huang(2017)Gao and Huang]{gao2017detecting}
Gao, L. and Huang, R.
\newblock Detecting online hate speech using context aware models.
\newblock \emph{arXiv preprint arXiv:1710.07395}, 2017.

\bibitem[Gorwa et~al.(2020)Gorwa, Binns, and Katzenbach]{gorwa2020algorithmic}
Gorwa, R., Binns, R., and Katzenbach, C.
\newblock Algorithmic content moderation: Technical and political challenges in
  the automation of platform governance.
\newblock \emph{Big Data \& Society}, 7\penalty0 (1):\penalty0
  2053951719897945, 2020.

\bibitem[Griesbach et~al.(2019)Griesbach, Reich, Elliott-Negri, and
  Milkman]{griesbach2019algorithmic}
Griesbach, K., Reich, A., Elliott-Negri, L., and Milkman, R.
\newblock Algorithmic control in platform food delivery work.
\newblock \emph{Socius}, 5:\penalty0 2378023119870041, 2019.

\bibitem[Hameleers \& van~der Meer(2020)Hameleers and van~der
  Meer]{hameleers2020misinformation}
Hameleers, M. and van~der Meer, T.~G.
\newblock Misinformation and polarization in a high-choice media environment:
  How effective are political fact-checkers?
\newblock \emph{Communication Research}, 47\penalty0 (2):\penalty0 227--250,
  2020.

\bibitem[Harrington(2010)]{harrington2010pop}
Harrington, B.
\newblock \emph{Pop finance: Investment clubs and the new investor populism}.
\newblock Princeton University Press, 2010.

\bibitem[Herring(2009)]{herring2009does}
Herring, C.
\newblock Does diversity pay?: Race, gender, and the business case for
  diversity.
\newblock \emph{American sociological review}, 74\penalty0 (2):\penalty0
  208--224, 2009.

\bibitem[Jehn et~al.(1999)Jehn, Northcraft, and Neale]{jehn1999differences}
Jehn, K.~A., Northcraft, G.~B., and Neale, M.~A.
\newblock Why differences make a difference: A field study of diversity,
  conflict and performance in workgroups.
\newblock \emph{Administrative science quarterly}, 44\penalty0 (4):\penalty0
  741--763, 1999.

\bibitem[Lacoste et~al.(2019)Lacoste, Luccioni, Schmidt, and
  Dandres]{lacoste2019quantifying}
Lacoste, A., Luccioni, A., Schmidt, V., and Dandres, T.
\newblock Quantifying the carbon emissions of machine learning.
\newblock \emph{arXiv preprint arXiv:1910.09700}, 2019.

\bibitem[Langvardt(2017)]{langvardt2017regulating}
Langvardt, K.
\newblock Regulating online content moderation.
\newblock \emph{Geo. LJ}, 106:\penalty0 1353, 2017.

\bibitem[Lewis(2018)]{lewis2018alternative}
Lewis, R.
\newblock Alternative influence: Broadcasting the reactionary right on youtube.
\newblock \emph{Data \& Society}, 18, 2018.

\bibitem[Liao \& Fu(2014)Liao and Fu]{liao2014can}
Liao, Q.~V. and Fu, W.-T.
\newblock Can you hear me now? mitigating the echo chamber effect by source
  position indicators.
\newblock In \emph{Proceedings of the 17th ACM conference on Computer supported
  cooperative work \& social computing}, pp.\  184--196, 2014.

\bibitem[Lindgren \& Holmstr{\"o}m(2020)Lindgren and
  Holmstr{\"o}m]{lindgren2020social}
Lindgren, S. and Holmstr{\"o}m, J.
\newblock A social science perspective on artificial intelligence: Building
  blocks for a research agenda.
\newblock \emph{Journal of Digital Social Research}, 2\penalty0 (3):\penalty0
  1--15, 2020.

\bibitem[Moss \& Metcalf(2020)Moss and Metcalf]{moss2020ethics}
Moss, E. and Metcalf, J.
\newblock Ethics owners: A new model of organizational responsibility in
  data-driven technology companies.
\newblock 2020.

\bibitem[Muller et~al.(2019)Muller, Fussell, Gao, Hinds, Oliveira, Reinecke,
  Robert~Jr, Siangliulue, Wulf, and Yuan]{muller2019learning}
Muller, M., Fussell, S.~R., Gao, G., Hinds, P.~J., Oliveira, N., Reinecke, K.,
  Robert~Jr, L., Siangliulue, K., Wulf, V., and Yuan, C.-W.
\newblock Learning from team and group diversity: Nurturing and benefiting from
  our heterogeneity.
\newblock In \emph{Conference Companion Publication of the 2019 on Computer
  Supported Cooperative Work and Social Computing}, pp.\  498--505, 2019.

\bibitem[Page(2008)]{page2008difference}
Page, S.~E.
\newblock \emph{The difference: How the power of diversity creates better
  groups, firms, schools, and societies-new edition}.
\newblock Princeton University Press, 2008.

\bibitem[Page(2019)]{page2019diversity}
Page, S.~E.
\newblock \emph{The diversity bonus: How great teams pay off in the knowledge
  economy}.
\newblock Princeton University Press, 2019.

\bibitem[Patton et~al.(2014)Patton, Hong, Ranney, Patel, Kelley, Eschmann, and
  Washington]{patton2014social}
Patton, D.~U., Hong, J.~S., Ranney, M., Patel, S., Kelley, C., Eschmann, R.,
  and Washington, T.
\newblock Social media as a vector for youth violence: A review of the
  literature.
\newblock \emph{Computers in Human Behavior}, 35:\penalty0 548--553, 2014.

\bibitem[Ravenelle(2019)]{ravenelle2019hustle}
Ravenelle, A.~J.
\newblock \emph{Hustle and gig: Struggling and surviving in the sharing
  economy}.
\newblock Univ of California Press, 2019.

\bibitem[Roberts(2019)]{roberts2019behind}
Roberts, S.~T.
\newblock \emph{Behind the screen: Content moderation in the shadows of social
  media}.
\newblock Yale University Press, 2019.

\bibitem[Sablosky(2021)]{sablosky2021dangerous}
Sablosky, J.
\newblock Dangerous organizations: Facebook’s content moderation decisions
  and ethnic visibility in myanmar.
\newblock \emph{Media, Culture \& Society}, pp.\  0163443720987751, 2021.

\bibitem[Schradie(2019)]{schradie2019revolution}
Schradie, J.
\newblock \emph{The revolution that wasn’t: How digital activism favors
  conservatives}.
\newblock Harvard University Press, 2019.

\bibitem[Shestakofsky(2017)]{shestakofsky2017working}
Shestakofsky, B.
\newblock Working algorithms: Software automation and the future of work.
\newblock \emph{Work and Occupations}, 44\penalty0 (4):\penalty0 376--423,
  2017.

\bibitem[Sheth et~al.(2011)Sheth, Bell, Arora, and Kaiser]{sheth2011towards}
Sheth, S.~K., Bell, J.~S., Arora, N., and Kaiser, G.~E.
\newblock Towards diversity in recommendations using social networks.
\newblock 2011.

\bibitem[Siddiquee(2020)]{siddiquee2020portrayal}
Siddiquee, M.~A.
\newblock The portrayal of the rohingya genocide and refugee crisis in the age
  of post-truth politics.
\newblock \emph{Asian Journal of Comparative Politics}, 5\penalty0
  (2):\penalty0 89--103, 2020.

\bibitem[Sinek(2009)]{sinek2009start}
Sinek, S.
\newblock \emph{Start with why: How great leaders inspire everyone to take
  action}.
\newblock Penguin, 2009.

\bibitem[Sloane \& Moss(2019)Sloane and Moss]{sloane2019ai}
Sloane, M. and Moss, E.
\newblock Ai’s social sciences deficit.
\newblock \emph{Nature Machine Intelligence}, 1\penalty0 (8):\penalty0
  330--331, 2019.

\bibitem[Sujarwoto et~al.(2019)Sujarwoto, Tampubolon, and
  Pierewan]{sujarwoto2019tool}
Sujarwoto, S., Tampubolon, G., and Pierewan, A.~C.
\newblock A tool to help or harm? online social media use and adult mental
  health in indonesia.
\newblock \emph{International journal of mental health and addiction},
  17\penalty0 (4):\penalty0 1076--1093, 2019.

\bibitem[Sylvia~Chou et~al.(2020)Sylvia~Chou, Gaysynsky, and
  Cappella]{sylvia2020we}
Sylvia~Chou, W.-Y., Gaysynsky, A., and Cappella, J.~N.
\newblock Where we go from here: health misinformation on social media, 2020.

\bibitem[Vijayaraghavan et~al.(2019)Vijayaraghavan, Larochelle, and
  Roy]{vijayaraghavan2019interpretable}
Vijayaraghavan, P., Larochelle, H., and Roy, D.
\newblock Interpretable multi-modal hate speech detection.
\newblock In \emph{AI for Social Good Workshop at the International Conference
  on Machine Learning}, 2019.

\bibitem[Wohn(2019)]{wohn2019volunteer}
Wohn, D.~Y.
\newblock Volunteer moderators in twitch micro communities: How they get
  involved, the roles they play, and the emotional labor they experience.
\newblock In \emph{Proceedings of the 2019 CHI conference on human factors in
  computing systems}, pp.\  1--13, 2019.

\bibitem[Wu(2017)]{wu2017attention}
Wu, T.
\newblock \emph{The attention merchants: The epic scramble to get inside our
  heads}.
\newblock Vintage, 2017.

\bibitem[Zuckerman \& Gessen(2019)Zuckerman and Gessen]{zuckerman2019unreality}
Zuckerman, E. and Gessen, M.
\newblock Unreality and social corrosion: Masha gessen and ethan zuckerman in
  conversation.
\newblock \emph{Journal of Design and Science}, \penalty0 (6), 2019.

\end{thebibliography}
\bibliographystyle{icml2021}

\end{document}